\documentclass[aps,prl,reprint,groupedaddress]{revtex4-2}
\UseRawInputEncoding
\usepackage[colorlinks,linkcolor=blue,urlcolor=blue,anchorcolor=blue,citecolor=blue]{hyperref}
\usepackage{amsmath}
\usepackage{graphicx}
\usepackage{dcolumn}
\usepackage{mathrsfs}
\usepackage{amssymb}
\usepackage{amsfonts,multirow}
\usepackage{bm}
\usepackage{color}
\usepackage{ulem} 
\usepackage{changes}
\setauthormarkup{}  
\setaddedmarkup{\textcolor{blue}{#1}}
\setdeletedmarkup{\textcolor{red}{\sout{#1}}}

\makeatletter
\colorlet{Changes@ColorDeleted}{red}
\makeatother

\begin{document}
	\draft
	\title{Experimental Observation of Non-Markovian Quantum Exceptional Points}
	\author{Hao-Long Zhang$^{1}$}
	\thanks{E-mail: These authors contribute equally to this work.}
	\author{Pei-Rong Han$^{2}$}
	\thanks{E-mail: These authors contribute equally to this work.}
	\author{Fan Wu$^{1}$}
	\author{Wen Ning$^{1}$}
	\thanks{E-mail: ningw@fzu.edu.cn}
	\author{Zhen-Biao Yang$^{1,3}$}
	\thanks{E-mail: zbyang@fzu.edu.cn}
	\author{Shi-Biao Zheng$^{1,3}$}
	\thanks{E-mail: t96034@fzu.edu.cn}
	\address{$^{1}$Fujian Key Laboratory of Quantum Information and Quantum\\
		Optics, College of Physics and Information Engineering, Fuzhou University,
		Fuzhou 350108, China\\
		$^{2}$School of Physics and Mechanical and Electrical Engineering, Longyan
		University, Longyan 364012, China\\
		$^{3}$Hefei National Laboratory, Hefei 230088, China}
	\date{\today }

\begin{abstract}
One of the most remarkable features that distinguish open systems from closed ones is the presence of exceptional points (EPs), where two or more eigenvectors of a non-Hermitian operator coalesce, accompanying the convergence of the corresponding eigenvalues. So far, EPs have been demonstrated on a number of platforms, ranging from classical optical systems to fully quantum-mechanical spin-boson models. In these demonstrations, the reservoir that induced the non-Hermiticity was treated as a Markovian one, without considering its memory effect. We here present the first experimental demonstration of non-Markovian quantum EPs, engineered by coupling a Josephson-junction-based qubit to a leaky electromagnetic resonator, which acts as a non-Markovian reservoir. We map out the spectrum of the extended Liouvillian superoperator by observing the quantum state evolution of the qubit and the pseudomode, in which the memory of the reservoir is encoded. We identify a twofold second-order EP and a third-order EP in the Liouvillian spectrum, which cannot be realized with a Markovian reservoir. Our results pave the way for experimental exploration of exotic phenomena associated with non-Markovian quantum EPs.
\end{abstract}

\maketitle
\vskip0.5cm

\narrowtext

An open quantum system can display exotic behaviors that are otherwise inaccessible. The environment of a quantum system can be modeled as a reservoir of electromagnetic modes, whose effects on the system depend upon its spectral structure \cite{Scully_Zubairy_1997}. When the spectrum is flat, the reservoir is memoryless and the system-reservoir interaction can be well described as a Markovian dynamics. In this case the system evolution is determined by a master equation, featuring the competition between the Hermitian Hamiltonian dynamics
and dissipation, described by a Lindblad dissipator. The dissipator involves
two competing processes, the coherent dissipation and quantum jump. The
coherent dissipation can be modeled as a non-Hermitian (NH) term, added to
the Hamiltonian. The competition between the Hermitian and NH terms gives
arise to the emergence of Hamiltonian exceptional points (HEPs), where two
or more eigenenergies become degenerate and the corresponding eigenvectors
coalesce \cite{Advance_in_Physics.69.249, RevModPhys.93.015005, NatRevPhys.4.745, Science.363.eaar7709}. The HEPs can lead to intriguing NH effects that are absent
in Hermitian systems, such as spectral real-to-complex transitions and
exceptional topology. So far, HEPs can be engineered in classical \cite{Science.363.eaar7709, NatMater.18.783,PhysRevLett.86.787,Nature.537.76,Nature.537.80,Science.359.1009,NatPhoton.13.623,Science.370.1077,PhysRevLett.129.084301,Nature.526.554,NatCommun.8.1368,SciAdv.7.eabj8905}, semiclassical \cite{NatCommun.10.855, Science.364.6443,NatPhys.18.385,PhysRevLett.126.083604,NatPhys.15.1232,ChinPhysB.30.100309}, and fully quantum-mechanical systems \cite{PhysRevLett.104.153601, PhysRevLett.131.260201,Arxiv.2407.00903,NatCommun.15.10293}.

In recent years, there has been growing interest in the formulation of exceptional points (EPs) based on the Liouvillian superoperator, which combines the effect of the quantum jumps with the NH Hamiltonian dynamics \cite{PhysRevA.100.062131,PhysRevA.101.013812,PhysRevA.101.062112,PhysRevA.102.033715,PRXQuantum.2.040346,AAPPSBull.34.22,PhysRevLett.127.140504,PhysRevLett.128.110402,NewJPhys.26.123032,NatCommun.13.6225,PhysRevLett.130.110402}. Such EPs, defined as
the degeneracies of the eigenvalues of the Liouvillian superoperator in
matrix representation, are referred to as Liouvillian EPs (LEPs). So far, the LEPs have been observed by coupling a qubit to an
artificially engineered Markovian reservoir \cite{PhysRevLett.127.140504, PhysRevLett.128.110402, NewJPhys.26.123032, NatCommun.13.6225, PhysRevLett.130.110402}. In a very recent theoretical work, the concept of LEPs was extended to the non-Markovian regime \cite{NatCommun.16.1289}. In this approach, the memory effects of the non-Markovian reservoir are captured by introducing an auxiliary bosonic mode, referred to as pseudomode (PM). The system-PMs dynamics is governed by an extended Liouvillian superoperator, which involves the degrees of freedom of both the system and PM. The incorporation of the PM makes the non-Markovian EPs fundamentally different from the Markovian EPs.

We here present the first experimental observation of such non-Markovian LEPs, engineered in a superconducting circuit, where an Xmon qubit is controllably coupled to its readout resonator. The resonator, holding a continuum of bosonic modes, serves as a structured reservoir for the qubit. The resulting non-Markovian dynamics is governed by an extended Liouvillian superoperator, which incorporates the state of the qubit with that of a PM, and can be expressed as a 9 $\times$ 9 NH matrix. The eigenvalues of this superoperator are extracted from the output joint density matrices of the system and PM, reconstructed for different interaction times. The parameter-space degenerate point is a combination of a third-order EP (EP3) and a twofold second-order EP (EP2). Such a purely non-Markovian quantum effect has not been observed so far.

\begin{figure*}
	\includegraphics{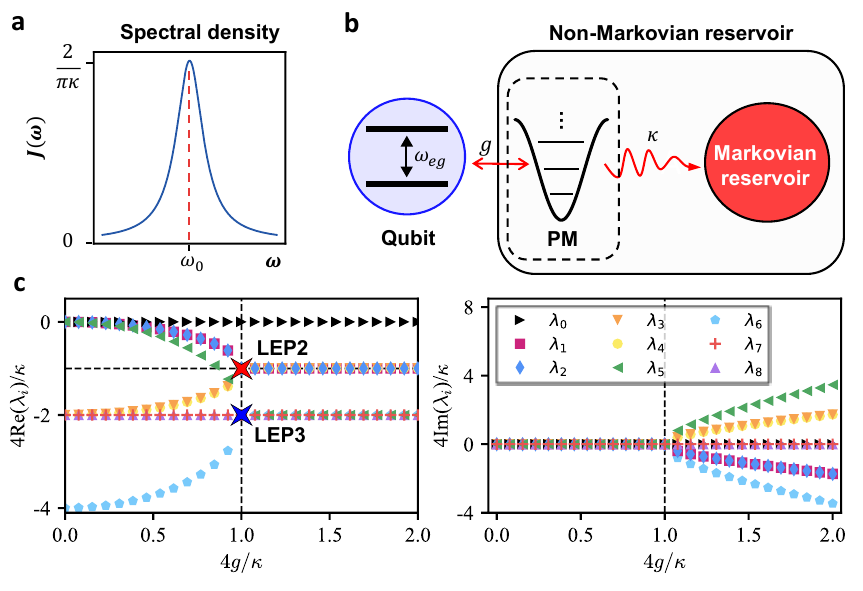} 
	\caption{(a) Spectral density $J(\omega )$ of the reservoir.
		The reservoir is a continuum of bosonic modes, having a Lorentzian shape with the spectral width, centered at the qubit frequency $\omega _{0}$. (b) Effective qubit-reservoir interaction model. The memory effect of the reservoir can be captured by a pseudomode (PM), which coherently swaps excitations with the qubit, and undergoes a continuous energy decay with a rate $\kappa$. (c) Real and imaginary parts of the spectrum of the extended Liouvillian superoperator. LEP2 and LEP3 are represented by red and blue stars, respectively. The LEP2 features simultaneous coalescence of the eigenvectors ${\bf V}_{1}$ with ${\bf V}_{3}$, and ${\bf V}_{2}$ with ${\bf V}_{4}$, while the LEP3 corresponds to coalescence of three eigenvectors ${\bf V}_{5}$, ${\bf V}_{6}$, and ${\bf V}_{7}$.}
	\label{f1}
\end{figure*}
We first give a brief introduction of the theoretical model, where a qubit
is coupled to a reservoir composed of a continuum of bosonic modes. In the
interaction picture, the qubit-reservoir dynamics is governed by the
Hamiltonian (setting $\hbar=1$)
\begin{equation}
\label{eq1}
H=\int_{0}^{\infty }{\rm d}\omega J(\omega )g(\omega )e^{i\delta (\omega
)t}a^{\dagger }(\omega )\left\vert l\right\rangle \left\langle u\right\vert
+\text{H.c.},
\end{equation}
where $J(\omega )$ is the spectral density of the reservoir, $a^{\dagger
}(\omega )$ is the creation operator for the bosonic mode with the frequency 
$\omega $, and $g(\omega )$\ and $\delta (\omega )$ are the
coupling strength and detuning between this bosonic mode and the qubit, whose upper and lower levels are respectively denoted as $\left\vert u\right\rangle $\ and $\left\vert l\right\rangle $. We here consider the Lorentzian spectral density \cite{RevModPhys.88.021002} [Fig.~\ref{f1}(a)],
\begin{equation}
J(\omega )=\frac{1}{\pi }\frac{\kappa /2}{(\omega -\omega _{0})^{2}+(\kappa
/2)^{2}},
\end{equation}
\begin{figure}
	\includegraphics[width=3.3in]{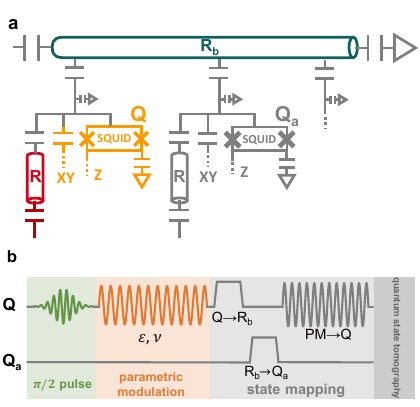} 
	\caption{(a) Sketch of the experimental system. The LEPs are realized with a circuit quantum electrodynamics architecture, where a bus
		resonator (R$_{b}$) connects five frequency-tunable Xmon qubits, one of
		which (Q) is used to test the non-Markovian dynamics. The readout resonator
		of Q (R) acts as a reservoir. (b) Pulse sequence. The experiment starts with
		the application of a $\pi /2$ pulse to Q, preparing it in the superposition
		state $(\left\vert l\right\rangle +i\left\vert u\right\rangle )/\sqrt{2}$.
		Then a parametric modulation with the frequency $\nu $ and amplitude $%
		\varepsilon $ is applied to Q, coupling it to R at a sideband. The effective
		coupling strength is controlled by $\varepsilon $. After a preset
		interaction time, the Q-R coupling is switched off. The output state of Q
		and the PM is read out by subsequently performing the state mappings: Q$%
		\rightarrow $R$_{b}\rightarrow $Q$_{a}$ and the PM$\rightarrow $Q.}
	\label{f2}
\end{figure}
where $\kappa $ represents the spectral width. Suppose that the spectral central frequency $\omega _{0}$ coincides with the qubit frequency. Because of the finite spectral width, the reservoir has a memory effect, which can be encoded in the dynamics of an effective damping mode, referred to as PM \cite{NatCommun.16.1289}, as illustrated by Fig.~\ref{f1}(b). Once the excitation is transferred from the qubit to the PM, it can be either transferred back to the qubit or leaked to the environment. The non-Markovian reservoir can be modeled as a PM immersed in a Markovian reservoir \cite{Scully_Zubairy_1997}, where the PM mode has a decaying rate equal to the spectral width of the Lorentzian reservoir. The evolution of the entire qubit-PM system is described by the master equation,
\begin{figure*}
	\includegraphics[width=7in]{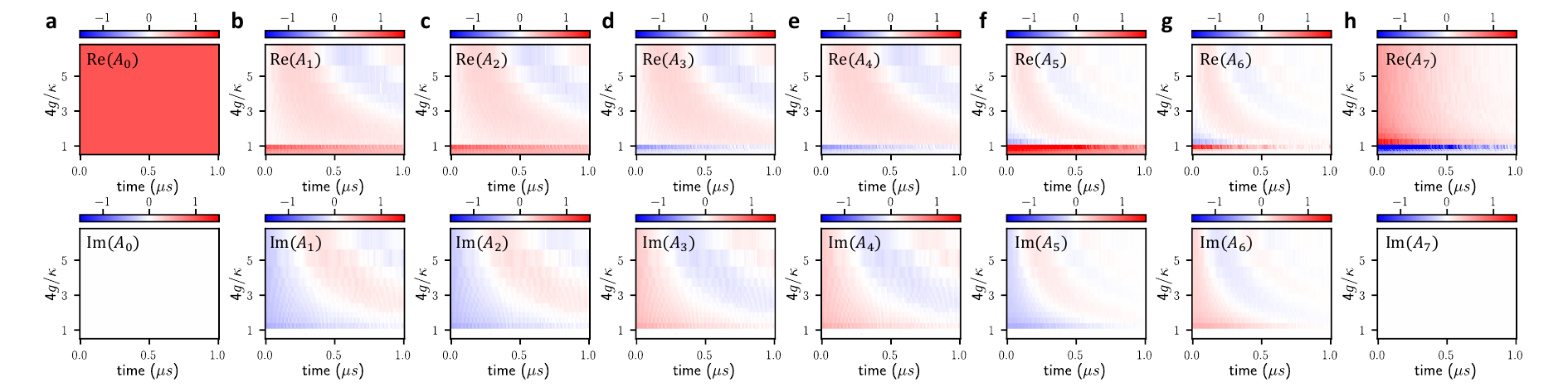} 
	\caption{Evolutions of amplitudes of the extended Liouvillian eigenvectors measured for different Q-PM coupling. The top row is Re$(A_j)$ and bottom row is Im$(A_j)$ (for $j=0,1,..,7$). The amplitudes for each time are obtained by expressing the reconstructed Q-PM density matrix in terms of the Liouvillian eigenvectors. The initial state does not include the last eigenvector, whose amplitude ($A_8$) remains zero and is not shown. }
	\label{f3}
\end{figure*}
\begin{equation}
\label{eq3}
\frac{d\rho }{dt}=-i\left(H_{\text{S,PM}}\rho-\rho H_{\text{S,PM}}^{\dagger} \right)+\kappa b\rho b^{\dagger },
\end{equation}
where $b^{\dagger }$ and $b$ denote the creation and annihilation operators
for the PM, and $H_{\text{S,PM}}$ is the NH Hamiltonian, given by 
\begin{equation}
H_{\text{S,PM}}=gb^{\dagger }\left\vert l\right\rangle \left\langle u\right\vert
+gb\left\vert u\right\rangle \left\langle l\right\vert -\frac{i\kappa }{2}
b^{\dagger }b,
\end{equation}
where $g$ is the coupling strength at the central frequency $\omega_0$, $g=g(\omega_0)$. The first two terms of $H_{\text{S,PM}}$ describe the reversible qubit-PM
swapping dynamics, while the last term accounts for the coherent nonunitary dissipation. The last term of the master equation $(\ref{eq3})$
describes the random quantum jump, by which the PM suddenly loses a photon.
The Hamiltonian dynamics and quantum jump can be incorporated into the
extended Liouvillian superoperator ${\cal L}_{\text{Q,PM}}$ with $d\rho/dt={\cal L}_{\rm Q,PM}\rho$.
Suppose that the reservoir is initially in the vacuum state. Then qubit-PM dynamics is restricted in the subspace with no more than one excitation, \{$
\left\vert l,0\right\rangle ,\left\vert u,0\right\rangle ,\left\vert l,1\right\rangle $\}, where the number in each ket denotes the photon number of the PM. The corresponding extended Liouvillian superoperator can be expressed as a $9\times 9$ NH matrix \cite{NatCommun.16.1289}, denoted as ${\cal L}_{\rm Q,PM}^{\rm matrix}$ \cite{PRLsupplement} (see Supplemental Material for details). The corresponding secular equation is
\begin{equation}
\mathcal{L}_{\text{Q,PM}}^{\text{matrix}} \mathbf{V}_j = \lambda_j \mathbf{V}_j.
\end{equation}
This matrix possesses three nondegenerate eigenvectors, ${\bf V}_{0}$, ${\bf V}_{5}$, and ${\bf V}_{6}$, and three pairs of degenerate eigenvectors (${\bf V} _{1};{\bf V} _{2}$), (${\bf V}_{3};{\bf V}_{4}$), and (${\bf V}_{7};{\bf V}_{8}$), which are detailed in Supplemental Material Sec.~S1 \cite{PRLsupplement}. The corresponding eigenvalues are given by
\begin{equation}
\begin{aligned}
\lambda _{0} &=0, \\
\lambda _{1} &=\lambda _{2}=-\kappa/4+\xi/2, \\
\lambda _{3} &=\lambda _{4}=-\kappa/4-\xi/2, \\
\lambda _{5} &=-\kappa/2+\xi, \\
\lambda _{6} &=-\kappa/2-\xi, \\
\lambda _{7} &=\lambda _{8}=-\kappa/2,
\end{aligned}
\end{equation}
where $\xi = \frac{1}{2}\sqrt{\kappa^2-16g^2}$. These eigenvalues versus the coupling strength are displayed in Fig.\ref{f1}(c). At the point $g=\kappa /4$, the spectrum display a second-order LEP (LEP2) and a third-order LEP (LEP3). We note that the LEP2 corresponds to the coalescence of the eigenvectors ${\bf V}_{1}$ and ${\bf V}_{3}$, as well as of ${\bf V}_{2}$ and ${\bf V}_{4}$, with the eigenvalue $-\kappa /4$. Such an EP essentially is a twofold LEP2. The LEP3 corresponds to the coalescence
of three eigenvectors ${\bf V}_{5}$, ${\bf V}_{6}$, and ${\bf V}_{7}$ with the eigenvalue $-\kappa/2 $. At the LEP3, ${\bf V}_{8}$ has the same eigenvalue as these three eigenvectors, but does not coalesce with them. These exotic spectral features originate from purely non-Markovian quantum effects, unaccessible in the Markovian reservoir, where the qubit undergoes a pure decay without energy and information backflow. 

The experiment is performed with a superconducting circuit, which involves a
bus resonator (R$_{b}$) and 5 frequency-tunable Xmon qubits, each
individually connected to a readout resonator, as sketched in Fig.~\ref{f2}(a) {\cite{PhysRevLett.131.260201}}. The
EPs are constructed with one of these qubits (Q), together with its readout resonator (R), which serves as a non-Markovian reservoir. The qubit has an energy relaxation time $T_{1} \approx  18.2$ $\rm{\mu} s$ and a Ramsey dephasing time $T_{2}^* \approx  6.3$ $\rm{\mu} s$, which are measured at its idle frequency $\omega_{\rm id,1}/2\pi \approx 6.01$ GHz. The spectrum of R exhibits a Lorentzian distribution with a spectral width of about $ 4.7$ MHz. A second qubit (Q$_{a}$) serves as an ancilla for mapping out the state of the PM, with the assistance of R$_{b}$ with a fixed frequency $5.582$ GHz and lifetime 13 $\mu$s. $T_{1}$ and $T_{2}^*$ for this qubit, measured at its idle frequency $\omega_{\rm id,2}/2\pi \approx5.46$ GHz, are 22.0 and 1.0 $\rm{\mu} s$, respectively. We note a recent superconducting circuit experiment \cite{PhysRevLett.132.200401}, where an ancilla transmon qubit was used as a reservoir for the test qubit, whose non-Markovianity was engineered by introducing thermal photons into its readout resonator. In distinct contrast, in the present experiment the readout resonator of the test qubit serves as its reservoir, with the non-Markovianity being controlled by their effective coupling strength \cite{ApplPhysLett.125.124003}.

The readout resonator can be modeled as a single photonic mode that is coupled to a vacuum Markovian reservoir \cite{Scully_Zubairy_1997}. Such a photonic mode corresponds to the PM with the decaying rate $\kappa=4.7$ MHz. The energy and information transferred from the qubit to the resonator can partially flow back to the qubit, as a consequence of the memory effect of such a non-Markovian reservoir.

Before the experiment, both qubits respectively stay at their idle frequencies $\omega_{\rm id,1/2}$, where they are effectively decoupled from each other and from the bus resonator due to the large detunings. The experiment starts with the application of a $\pi/2$ pulse to Q at its idle frequency $\omega_{\rm id,1}$, transforming it from the lower state $\vert l \rangle$ to the superposition, $(\left\vert l\right\rangle +i\left\vert u\right\rangle )/%
\sqrt{2}$. The pulse sequence is shown in Fig.~\ref{f2}(b). The central frequency of the Lorentzian spectrum in R is $\omega_0/2\pi \approx 6.66$ GHz, which is much higher than Q's maximum frequency. To couple Q to R, a 
parametric modulation with the adjustable frequency $\nu $ and amplitude $%
\varepsilon $ is applied to Q, mediating the Q-R swapping interaction at one
sideband, as detailed in \cite{physRevA.86.022305, PhysRevB.87.220505, PhysRevLett.131.260201, NatCommun.12.5924, PRXQuantum.5.020321}. The induced sideband Q-R coupling is described by the Hamiltonian of Eq. (\ref{eq1}). The Q-PM dynamics is governed by the master equation of Eq. (\ref{eq3}), with the effective interaction strength adjustable by $\varepsilon$. 

After a preset interaction time, the parametric modulation is switched off, so that Q is decoupled from R. To read out the joint Q-PM state, the quantum state of Q is transferred to the ancilla Q$_{a}$ through R$_{b}$, following which PM's state is mapped to Q through the sideband interaction. As the PM contains no more than one photon, it can be considered a qubit and the PM-Q quantum state transfer is realized by the swapping gate. Then the ${\rm Q}_{a}$-Q density matrix is reconstructed by quantum state tomography. With a suitable correction, the measured Q$_{a}$-Q state corresponds to the Q-R state just before the mapping operations. It is impossible to determine whether the component $\left\vert l,0\right\rangle $ originates from the initial population or from the quantum jump $\left\vert l,1\right\rangle \rightarrow \left\vert l,0\right\rangle $, and consequently, the state trajectory without quantum jump cannot be postselected. This is in distinct contrast with the previous experiment \cite{PhysRevLett.131.260201}, where coherent dynamics associated with the NH Hamiltonian is confined in the single-excitation subspace. As the Liouvillian superoperator enables the effect of quantum jumps to be incorporated into the description of the dynamical process, the associated spectrum can display much richer and more exotic EPs than the NH Hamiltonian eigenspectrum.
To obtain the evolutions of the amplitudes associated with different eigenvectors, we expand the measured density matrix in terms of the eigenvectors of the extended Liouvillian superoperator:
\begin{equation}
\mathbf{V}(t) = e^{\mathcal{L}_{\text{Q,PM}}t}\mathbf{V}(0)=\sum_{j=0}^{8} A_j(t)\mathbf{V}_j,
\end{equation}
where $\mathbf{V}(t)$ is the vector corresponding to the output qubit-PM density matrix at time $t$ and $A_j(t) = A_j(0)e^{\lambda_j t}$ is the amplitude associated with $\mathbf{V}_j$. Fig.~\ref{f3}(a)-(h) display the measured amplitudes ($A_{j}$) of the first eight extended Liouvillian eigenvectors versus the interaction time and effective coupling. As expected, the amplitude $A_{0}$ remains unchanged. Both the real and imaginary parts of $A_{7}$ display monotonous decaying behaviors for all values of the coupling strength, while those of $A_{j}$ ($j=1,2,...,6$) also monotonously decays when $g<\kappa /4$, but exhibit local oscillation features for $g>\kappa /4$, indicating a spectral transition. After a long time evolution, the system tends to the steady state $\left\vert g,0\right\rangle $, so that only the amplitude $A_{0}$ survives. 
\begin{figure*}
	\includegraphics[width=0.9\linewidth]{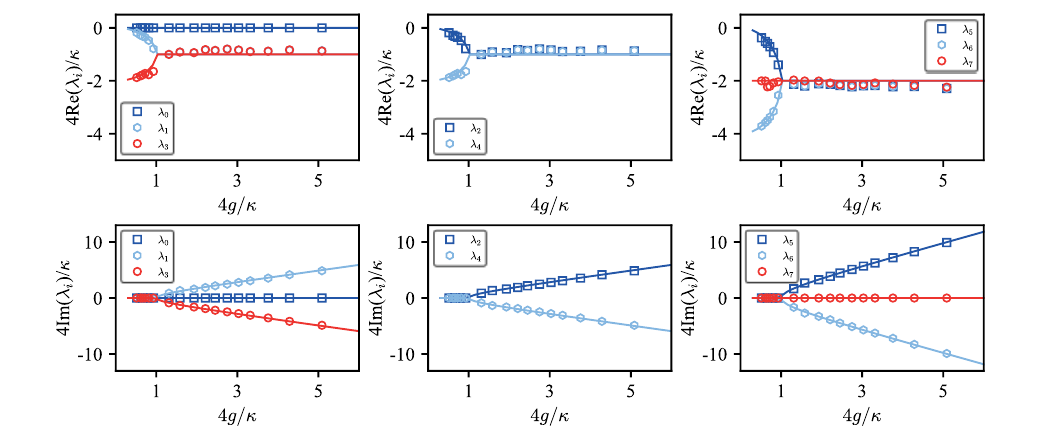} 
	\caption{Reconstructed Liouvillian spectrum. The eigenvalues associated with the first eight eigenvectors of the extended Liouvillian superoperator are inferred by the exponential fitting of the time evolving amplitude $A_j(t) = A_j(0)e^{\lambda_jt}$.}
   \label{f4}
\end{figure*}

Through exponential fitting of the evolution of the amplitude associated to each eigenvector, we can infer the corresponding eigenvalue. The thus-obtained Liouvillian spectrum is displayed in Fig.~\ref{f4}. Within the range of the error, we have $\lambda _{1}=\lambda _{2}$ and $\lambda _{3}=\lambda _{4}$. Each of the two twofold degenerate eigenvalues undergoes a real-to-complex transition at $g=\kappa /4$, featuring a twofold LEP2. We note that such a twofold LEP2 is unobservable when there is no quantum coherence between the one-excitation state components and the ground state. $\lambda _{5}$ and $\lambda _{6}$ also exhibit a real-to-complex transition at $g=\kappa /4$, where they combine with $\lambda _{7}$. At this point, the eigenvectors ${\bf V}_{5}$, ${\bf V}_{6}$, and ${\bf V}_{7}$ also coalesce, indicating the emergence of an LEP3. Because of the degeneracy of the eigenvalues, the choice of the corresponding eigenvectors is not unique. The initial value of the related amplitude $A_j (t)$ depends on such a choice. However, the time-dependent factor $e^{\lambda_j t}$ is independent of this choice. The eigenvalue $\lambda_j$, extracted by the corresponding exponential fitting, does not depend on $A_j (0)$. We note that the Liouvillian spectrum can also be obtained by quantum process tomography which is realized with different input states, as demonstrated in Ref. \cite{NewJPhys.26.123032}. For the present method, we only use a single input state, but need to reconstruct the output density matrices for different evolution times in order to obtain the eigenvalues.

In conclusion, we have experimentally demonstrated non-Markovian quantum EPs in a superconducting circuit, where an Xmon qubit is controllably coupled to its readout resonator, which acts as a structured reservoir with a continuum of bosonic modes. In addition to inducing a decaying channel, such a reservoir can coherently couple the qubit levels, and promote the effective dimension of the system. These unique non-Markovian effects enable the simultaneous emergence of EPs of different orders at the same point of the parameter space. Our work opens the door to experimental investigations of NH phenomena that are inaccessible in Markovian open systems.

\textit{Acknowledgments}---This work was supported by the National Natural Science Foundation of China
(Grants No.~12474356, No.~12475015, No.~12274080, No.~12505016, No.~12505021), Quantum Science and Technology-National Science and Technology Major Project (Grant No.~2021ZD0300200) and the Natural Science Foundation of Fujian Province (Grants No.~2025J01383 and No.~2025J01465).

\textit{Data availability}---The data that support the findings of this article are not publicly available. The data are available from the authors upon reasonable request.
\bibliography{reference}

\end{document}


\title{Supplementary Materials for \\``Experimental observation of non-Markovian quantum exceptional points"}
	
\author{Hao-Long Zhang$^{1}$}
	\thanks{E-mail: These authors contribute equally to this work.}
	\author{Pei-Rong Han$^{2}$}
	\thanks{E-mail: These authors contribute equally to this work.}
	\author{Fan Wu$^{1}$}
	\author{Wen Ning$^{1}$}
	\thanks{E-mail: ningw@fzu.edu.cn}
	\author{Zhen-Biao Yang$^{1,3}$}
	\thanks{E-mail: zbyang@fzu.edu.cn}
	\author{Shi-Biao Zheng$^{1,3}$}
	\thanks{E-mail: t96034@fzu.edu.cn}
	\address{$^{1}$Fujian Key Laboratory of Quantum Information and Quantum\\
		Optics, College of Physics and Information Engineering, Fuzhou University,
		Fuzhou 350108, China\\
		$^{2}$School of Physics and Mechanical and Electrical Engineering, Longyan
		University, Longyan 364012, China\\
		$^{3}$Hefei National Laboratory, Hefei 230088, China}

\vskip0.5cm

\narrowtext

\maketitle
	
\tableofcontents

\section{Solution of the secular equation}

The dynamics of the entire Q-PM system can be described by the Lindblad master equation
\begin{equation}
\frac{\partial\rho}{\partial t} = -i [H,\rho] + \frac{\kappa}{2}(2 b \rho b^{\dagger}-\{b^{\dagger}b,\rho\}) \equiv \mathcal{L}_{\text{Q,PM}} \rho,
\end{equation}
where the system Hamiltonian is given by $H=g(b^{\dagger}|l\rangle\langle u|+b|u\rangle\langle l|)$, the system state $\rho$ is a $3\times3$ density matrix in the subspace $\{|l,0\rangle,|u,0\rangle,|l,1\rangle\}$, and the extended Liouvillian superoperator $\mathcal{L}_{\text{Q,PM}}$ in the defined matrix representation can be written as
\begin{equation}
\begin{aligned}
\mathcal{L}_{\text{Q,PM}}^{\text{matrix}} &= -i(H \otimes I - I \otimes H^{\top}) 
+ \frac{\kappa}{2} (2 b \otimes b^{*} -b^{\dagger}b \otimes I - I \otimes b^{\top}b^{*})\\
&= -i(H \otimes I - I \otimes H^{\top}) 
+ \frac{\kappa}{2} (2 b \otimes b -b^{\dagger}b \otimes I - I \otimes b^{\dagger}b).
\end{aligned}
\end{equation}
Here $\otimes$ is Kronecker product operation, $\top$ and $*$ represent the transpose and complex conjugate operations, respectively. Therefore, in the subspace spanned by $\{|l,0\rangle,|u,0\rangle,|l,1\rangle\}$, the matrix form of the extended Liouvillian superoperator is a $9 \times 9$ matrix:
\begin{equation}
\mathcal{L}_{\text{Q,PM}}^{\text{matrix}} =\left(
\begin{array}{ccccccccc}
0 & 0 & 0 & 0 & 0 & 0 & 0 & 0 & \kappa \\
0 & 0 & ig & 0 & 0 & 0 & 0 & 0 & 0 \\
0 & ig & -\frac{\kappa}{2} & 0 & 0 & 0 & 0 & 0 & 0 \\
0 & 0 & 0 & 0 & 0 & 0 & ig & 0 & 0 \\
0 & 0 & 0 & 0 & 0 & ig & 0 & -ig & 0 \\
0 & 0 & 0 & 0 & ig & -\frac{\kappa}{2} & 0 & 0 & -ig \\
0 & 0 & 0 & -ig & 0 & 0 & -\frac{\kappa}{2} & 0 & 0 \\
0 & 0 & 0 & 0 & -ig & 0 & 0 & -\frac{\kappa}{2} & ig \\
0 & 0 & 0 & 0 & 0 & -ig & 0 & ig & -\kappa \\

\end{array}
\right).
\end{equation}
By solving the secular equation
\begin{equation}
\mathcal{L}_{\text{Q,PM}}^{\text{matrix}} \mathbf{V}_j = \lambda_j \mathbf{V}_j,
\end{equation}
the eigenvalues $\lambda_j$ and the eigenvectors $\mathbf{V}_j$ (represented by the density matrix form $\rho_j$) of the extended Liouvillian superoperator can be obtained, written as follows: 
\begin{equation}
\lambda_0 = 0,\quad
\mathbf{V}_0 =
\left(
\begin{array}{cccccccccc}
1 & 0 & 0 &
0 & 0 & 0 &
0 & 0 & 0 
\end{array}
\right)^{\rm T};
\end{equation}

\begin{equation}
\lambda_1  = -\frac{\kappa}{4}+\frac{\xi}{2}, \quad
\mathbf{V}_1 =\frac{1}{N_1}
\left(
\begin{array}{cccccccccc}
0 & 0 & 0 &
i\frac{\kappa}{2}+i\xi  & 0 & 0 &
2g & 0 & 0
\end{array}
\right)^{\rm T},
\end{equation}

\begin{equation}
\lambda_2 = -\frac{\kappa}{4}+\frac{\xi}{2}, \quad
\mathbf{V}_2 =\frac{1}{N_2}
\left(
\begin{array}{cccccccccc}
0 & -i\frac{\kappa}{2}-i\xi  & 2g &
0 & 0 & 0 &
0 & 0 & 0
\end{array}
\right)^{\rm T},
\end{equation}
with
$
N_{1,2} = \left\{
\begin{aligned}
\nonumber
\sqrt{\kappa\left(\frac{\kappa}{2}+\xi\right)}, & \quad g<\frac{\kappa}{4} \\
2\sqrt{2}g, & \quad g\geq \frac{\kappa}{4}
\end{aligned}
\right.
$ and $\xi = \frac{1}{2}\sqrt{\kappa^2-16g^2}$;

\begin{equation}
\lambda_3  = -\frac{\kappa}{4}-\frac{\xi}{2}, \quad
\mathbf{V}_3 =\frac{1}{N_3}
\left(
\begin{array}{cccccccccc}
0 & 0 & 0 &
i\frac{\kappa}{2}-i\xi  & 0 & 0 &
2g & 0 & 0
\end{array}
\right)^{\rm T},
\end{equation}

\begin{equation}
\lambda_4 = -\frac{\kappa}{4}-\frac{\xi}{2}, \quad
\mathbf{V}_4 =\frac{1}{N_4}
\left(
\begin{array}{cccccccccc}
0 & -i\frac{\kappa}{2}+i\xi  & 2g &
0 & 0 & 0 &
0 & 0 & 0
\end{array}
\right)^{\rm T},
\end{equation}
with
$
N_{3,4} = \left\{
\begin{aligned}
\nonumber
\sqrt{\kappa\left(\frac{\kappa}{2}-\xi\right)}, & \quad g<\frac{\kappa}{4} \\
	2\sqrt{2}g, & \quad g\geq \frac{\kappa}{4}
	\end{aligned};\right.
$

\begin{equation}
\lambda_5 = -\frac{\kappa}{2}-\xi, \quad
\mathbf{V}_5 =\frac{1}{N_5}
\left(
\begin{array}{cccccccccc}
-\kappa & 0 & 0 &
0 & \frac{\kappa}{2}-\xi & i2g &
0 & -i2g & \frac{\kappa}{2}+\xi
\end{array}
\right)^{\rm T},
\end{equation}

\begin{equation}
\lambda_6 = -\frac{\kappa}{2}+\xi, \quad
\mathbf{V}_6 =\frac{1}{N_6}
\left(
\begin{array}{cccccccccc}
-\kappa & 0 & 0 &
0 & \frac{\kappa}{2}+\xi & i2g &
0 & -i2g & \frac{\kappa}{2}-\xi
\end{array}
\right)^{\rm T},
\end{equation}
with 
$
N_{5,6} = \left\{
\begin{aligned}
\nonumber
\sqrt{2}\kappa, & \quad g<\frac{\kappa}{4} \\
\sqrt{\kappa^2+16g^2}, & \quad g\geq \frac{\kappa}{4}
\end{aligned};\right.
$
\begin{equation}
\lambda_7 = -\frac{\kappa}{2}, \quad
\mathbf{V}_7 =\frac{1}{\sqrt{2\kappa^2+96g^2}}
\left(
\begin{array}{cccccccccc}
-8g & 0 & 0 &
0 & 4g & i\kappa &
0 & -i\kappa & 4g
\end{array}
\right)^{\rm T};
\end{equation}

\begin{equation}
\lambda_8 = -\frac{\kappa}{2}, \quad
\mathbf{V}_8 =\frac{1}{\sqrt{2}}
\left(
\begin{array}{cccccccccc}
0 & 0 & 0 &
0 & 0 & 1 &
0 & 1 & 0
\end{array}
\right)^{\rm T}.
\end{equation}

When $\xi=0$ with $g=\frac{\kappa}{4}$, the above eigenvalues $\lambda_j$ and eigenvectors $\rho_j$ are simplified as: 
\begin{equation}
\lambda_1=\lambda_3=-\frac{\kappa}{4}, \quad 
\mathbf{V}_1=\mathbf{V}_3=\frac{1}{\sqrt{2}}
\left(
\begin{array}{ccccccccc}
0 & 0 & 0 &
i& 0 & 0 &
1 & 0 & 0 
\end{array}
\right)^{\rm T};
\end{equation}
\begin{equation}
\lambda_2=\lambda_4=-\frac{\kappa}{4}, \quad 
\mathbf{V}_2=\mathbf{V}_4=\frac{1}{\sqrt{2}}
\left(
\begin{array}{ccccccccc}
0 & -i & 1 &
0 & 0 & 0 &
0 & 0 & 0 
\end{array}
\right)^{\rm T};
\end{equation}
\begin{equation}
\lambda_5=\lambda_6 =\lambda_7=-\frac{\kappa}{2}, \quad 
\mathbf{V}_5=\mathbf{V}_6=\mathbf{V}_7=\frac{1}{2\sqrt{2}}
\left(
\begin{array}{ccccccccc}
-2 & 0 & 0 &
 0 & 1 & i  &
0 & -i & 1 
\end{array}
\right)^{\rm T}.
\end{equation}

These results indicate that a two-fold LEP2 and a LEP3 simultaneously merge at the same point. As the eigenvalue $\lambda_7$ is always equal to $\lambda_8$, the choice of the corresponding eigenvectors $\mathbf{V}_7$ and $\mathbf{V}_8$ is not unique. However, with other choices the position of EPs is associated with more different eigenvectors than it is necessary.

The nine eigenvectors can be rewritten as the following 3×3 matrices,
\begin{equation}
\rho_0 =
\left(
\begin{array}{ccc}
1 & 0 & 0 \\
0 & 0 & 0 \\
0 & 0 & 0 
\end{array}
\right);
\end{equation}

\begin{equation}
\rho_1 =\frac{1}{N_1}
\left(
\begin{array}{ccc}
0 & 0 & 0 \\
i\frac{\kappa}{2}+i\xi  & 0 & 0 \\
2g & 0 & 0
\end{array}
\right),
\end{equation}

\begin{equation}
\rho_2 =\frac{1}{N_2}
\left(
\begin{array}{ccc}
0 & -i\frac{\kappa}{2}-i\xi  & 2g \\
0 & 0 & 0 \\
0 & 0 & 0
\end{array}
\right),
\end{equation}
with
$
N_{1,2} = \left\{
\begin{aligned}
\nonumber
\sqrt{\kappa\left(\frac{\kappa}{2}+\xi\right)}, & \quad g<\frac{\kappa}{4} \\
2\sqrt{2}g, & \quad g\geq \frac{\kappa}{4}
\end{aligned}
\right.
$ and $\xi = \frac{1}{2}\sqrt{\kappa^2-16g^2}$;

\begin{equation}
\rho_3 =\frac{1}{N_3}
\left(
\begin{array}{ccc}
0 & 0 & 0 \\
i\frac{\kappa}{2}-i\xi  & 0 & 0 \\
2g & 0 & 0
\end{array}
\right),
\end{equation}

\begin{equation}
\rho_4 =\frac{1}{N_4}
\left(
\begin{array}{ccc}
0 & -i\frac{\kappa}{2}+i\xi  & 2g \\
0 & 0 & 0 \\
0 & 0 & 0
\end{array}
\right),
\end{equation}
with
$
N_{3,4} = \left\{
\begin{aligned}
\nonumber
\sqrt{\kappa\left(\frac{\kappa}{2}-\xi\right)}, & \quad g<\frac{\kappa}{4} \\
	2\sqrt{2}g, & \quad g\geq \frac{\kappa}{4}
	\end{aligned};\right.
$

\begin{equation}
\rho_5 =\frac{1}{N_5}
\left(
\begin{array}{ccc}
-\kappa & 0 & 0 \\
0 & \frac{\kappa}{2}-\xi & i2g \\
0 & -i2g & \frac{\kappa}{2}+\xi
\end{array}
\right),
\end{equation}

\begin{equation}
\rho_6 =\frac{1}{N_6}
\left(
\begin{array}{ccc}
-\kappa & 0 & 0 \\
0 & \frac{\kappa}{2}+\xi & i2g \\
0 & -i2g & \frac{\kappa}{2}-\xi
\end{array}
\right),
\end{equation}
with 
$
N_{5,6} = \left\{
\begin{aligned}
\nonumber
\sqrt{2}\kappa, & \quad g<\frac{\kappa}{4} \\
\sqrt{\kappa^2+16g^2}, & \quad g\geq \frac{\kappa}{4}
\end{aligned};\right.
$
\begin{equation}
\rho_7 =\frac{1}{\sqrt{2\kappa^2+96g^2}}
\left(
\begin{array}{ccc}
-8g & 0 & 0 \\
0 & 4g & i\kappa \\
0 & -i\kappa & 4g
\end{array}
\right);
\end{equation}

\begin{equation}
\rho_8 =\frac{1}{\sqrt{2}}
\left(
\begin{array}{ccc}
0 & 0 & 0 \\
0 & 0 & 1 \\
0 & 1 & 0
\end{array}
\right).
\end{equation}
With this reformulation, the eigenvectors $\mathbf{V}_j$ with $j = 1$ to $8$ correspond to traceless, unphysical matrices $\rho_j$, while they, together with the steady-state eigenvector $\mathbf{V}_0$, form a basis, in terms of which the output Q-PM density matrix at any time can be expanded.

To realize EPs in a Markovian reservoir, it is necessary to coherently couple the qubit to an external field. For this case, the dynamics is restricted in a two-dimensional Hilbert space so that the Liouvillian superoperator corresponds to a $4\times 4$ NH matrix. Therefore, the non-Markovian reservoir plays a three-fold role in the formation of the Liouvillian spectral structure: producing a vacuum Rabi splitting, inducing dissipation and quantum jumps, and extending the Hilbert space of the qubit by entangling its degrees of freedom with the qubit.

\section{Evolution of the qubit-PM system}
With the extended Liouvillian superoperator $\mathcal{L}_{\text{Q,PM}}$, the evolution of the system state can be described by
\begin{equation}
\mathbf{V}(t)= e^{\mathcal{L}_{\text{Q,PM}}t}\mathbf{V}(0)=\sum_{j=0}^{8} A_j(t)\mathbf{V}_j,
\end{equation}
where $\mathbf{V}(t)$ is the density operator in the form of the superoperator and $A_j(t) = A_j(0)e^{\lambda_j t}$ is the amplitude associated with $\mathbf{V}_j$. For the initial state $|\psi(0)\rangle = \frac{1}{\sqrt{2}}\left(|g,0\rangle+
i|e,0\rangle\right)$, it gives the evolution as
\begin{equation}
\begin{aligned}
\mathbf{V}(t) =\quad& e^{\mathcal{L}_{\text{Q,PM}}t} \mathbf{V}(0) \\
=\quad& \mathbf{V}_0 - \frac{g^3}{(\xi)^2\lambda_5\lambda_6} N_7 e^{\lambda_7t}\mathbf{V}_7
- \frac{g^2}{2(\xi)^2\lambda_5} N_5 e^{\lambda_5t}\mathbf{V}_5
- \frac{g^2}{2(\xi)^2\lambda_6} N_6 e^{\lambda_6t}\mathbf{V}_6 \\
& - \frac{1}{4\xi} \left( N_3 e^{\lambda_3 t} \mathbf{V}_3 + N_4 e^{\lambda_4t}\mathbf{V}_4 - N_1 e^{\lambda_1t}\mathbf{V}_1 - N_2 e^{\lambda_2 t}\mathbf{V}_2 \right).
\end{aligned}
\label{rho(t)}
\end{equation}

The last eigenvector is not included in the expansion, so that its amplitude remains 0. As this eigenvector has the same eigenvalue as the eighth one, and does not coalesce with any other, its absence does not affect the identification of the LEPs in any way. In each subfigure of Fig.~3 in the main text, the upper and lower panels respectively denote the real and imaginary parts of the corresponding amplitude.
\begin{figure}
	\includegraphics[width=3.5in]{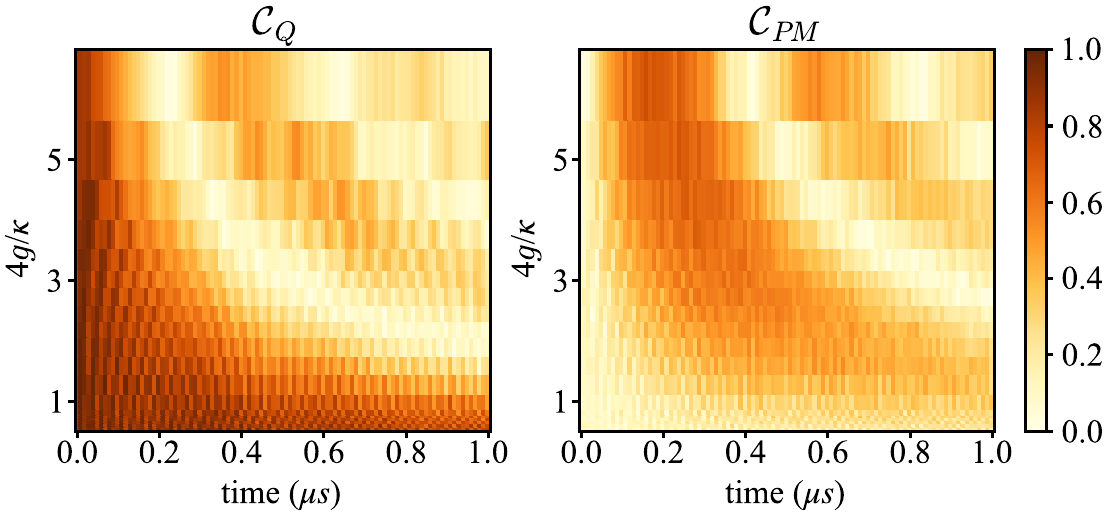} 
	\caption{Observed evolutions of the qubit and PM
		coherences. The qubit (PM) coherence is defined as the sum of the absolute
		values of the off-diagonal elements of the reduced density matrix for the
		qubit (PM).}
	\label{f5}
\end{figure}

To study the population dynamics of the qubit, the degrees of freedom of the pseudomode are traced out, i.e.,
\begin{equation}\label{eS19}
\rho_{\text{Q}}(t) = \text{tr}_{\text{PM}}\left[ \rho(t) \right] = \sum_{j=0}^{8} A_j(t)\rho_{\text{Q},j},
\end{equation}
with $\rho_{\text{Q},j} = \text{tr}_{\text{PM}}(\rho_j)$ being the reduced eigenmatrices. In the basis of $\{|u\rangle,|l\rangle\}$, $\rho_{\text{Q}}(t)$ in Eq. (\ref{eS19}) can be expressed as 
\begin{equation}
\rho_{\text{Q}}(t) =
\left(
\begin{array}{cc}
\rho_{uu}(t) & \rho_{ul}(t) \\
\rho_{lu}(t) & \rho_{ll}(t)
\end{array}
\right),
\end{equation}
where the four matrix components are detailed as 
\begin{equation}
\left\{
\begin{aligned}
\rho_{uu}(t) &= \frac{e^{-\frac{\kappa}{2} t}}{8(\xi)^2} \left[ 8g^2 + \left(8g^2-\kappa^2\right) \cos{(\xi t)} + 2\kappa (\xi)\sin{\xi t} \right] \\
\rho_{ll}(t) &= 1 - \rho_{uu}(t) \\
\rho_{ul}(t) &= \rho_{lu}^{*} (t)= \frac{i}{4\xi}e^{-\frac{\kappa}{4}t} \left[2\xi\cos{\left(\frac{\xi}{2}t\right)} + \kappa\sin{\left(\frac{\xi}{2}t\right)}\right]
\end{aligned}
\right..
\label{P(t)}
\end{equation}
From Eq.~(\ref{P(t)}), it is clear that the population exhibits an oscillatory behavior for $g>\kappa/4$, but shows an overdamped evolution for $g\leq\kappa/4$. 
\begin{figure*}
    \centering
    \includegraphics{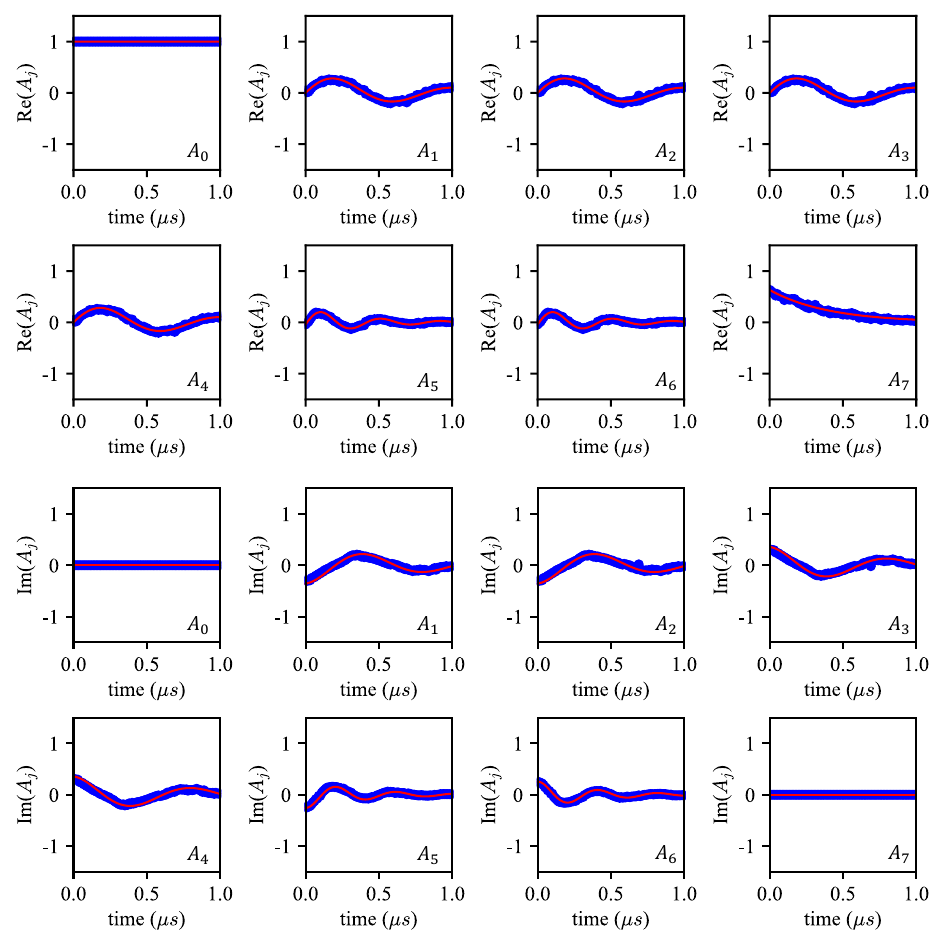}
	\caption{The real and imaginary parts of the amplitudes associated with the eigenmatrices $A_j(t)$ versus the evolution time $t$, for $g/2\pi \approx  1.242$ MHz. Dots and lines denote the experimentally measured and fitting results, respectively.}
    \label{fig1}
\end{figure*}

\section{Quantum coherences for the qubit and PM}
These exotic LEPs originate from the non-Markovianity of the reservoir,
which is manifested in the memory effect encoded in the PM. By participating in the coherent dynamics of the qubit, PM expands the Hilbert space to three dimensions. When the qubit is initially in the upper level, there are no quantum coherences between the zero- and one-excitation states, defined as ${\cal C}_{\rm Q}=2\left\vert \left\langle u,0\right\vert \rho \left\vert
l,0\right\rangle \right\vert $ and ${\cal C}_{\rm PM}=2\left\vert \left\langle
l,1\right\vert \rho \left\vert l,0\right\rangle \right\vert $, where $\rho $
is the joint qubit-PM density matrix. Here ${\cal C}_{\rm Q}$ and ${\cal C}_{\rm PM}$
correspond to the qubit coherence and PM coherence, respectively. For the state of Eq.~(\ref{rho(t)}), the corresponding quantum coherences are given by
\begin{equation}
\mathcal{C}_{\text{Q}}(t) = \left|\frac{i}{2\xi}\left[\frac{\kappa}{2}\left(e^{\lambda_3t}-e^{\lambda_1t}\right)-\xi\left(e^{\lambda_3t}+e^{\lambda_1t}\right)\right]\right| ,
\end{equation}
and
\begin{equation}
\mathcal{C}_{\text{PM}}(t) =\left|\frac{g}{\xi}\left(e^{\lambda_3t}-e^{\lambda_1t}\right)\right|.
\end{equation}
It can be verified that, at the point $g=\kappa/4$, the dynamics of $\mathcal{C}_{\text{Q}}$ and $\mathcal{C}_{\text{PM}}$ also show a transition from decaying to oscillatory behavior. Without these coherences, the Liouvillian expansion does not include the
eigenvectors ${\bf V}_{1}$, ${\bf V}_{2}$, ${\bf V}_{3}$, and ${\bf V}_{4}$,
which are responsible for the emergence of the two-fold LEP2. By preparing the qubit in the superposition state,
we can observe how these quantum coherences evolve. Fig.~\ref{f5} display the evolutions of ${\cal C}_{\rm Q}$ and ${\cal C}_{\rm PM}$ for different values of $g$. As expected, these coherences are varied with time $t$ no matter the control parameter is above or below the point $g = \kappa/4$. The two-fold LEP2 corresponds to the critical damping point for ${\cal C}_{\rm Q}$: It exhibits an underdamped (oscillatory) behavior above this point, but shows an overdamped evolution below this point.

\section{Extraction of eigenvalues}

As shown in Eq.~(\ref{rho(t)}), the state evolutions of the entire Q-PM system can be expanded with eigenmatrices and the associated amplitudes $A_j(t)$ vary exponentially in time $t$ with rate $\lambda_j$. In our experiment, the system state $\rho(t)$ at each moment can be reconstructed by the method of quantum state tomography. Through indentifying the information of all the components of the density matrix $\rho(t)$ at different times, and then with resorting to the least square method for fitting of $A_j(t)$ to fit the exponential function $C_j e^{B t}$, we can extract the best-fit eigenvalues, i.e., $B = \lambda_j$.
\begin{figure}
    \centering
    \includegraphics[width=1.0\textwidth]{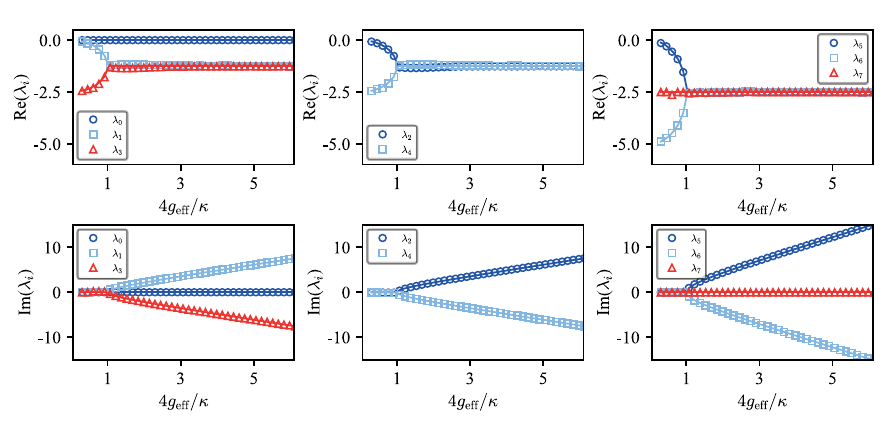}
    \caption{Numerically extracted eigenvalues of the Liouvillian superoperator versus the effective qubit-PM coupling strength. These eigenvalues are obtained by exponentially fitting the amplitudes associated with the corresponding eigenvectors, governed by the master equation, where the qubit-PM interaction is described by the full Hamiltonian of Eq.~(\ref{H_sideband}).}
    \label{fig2}
\end{figure}
\begin{figure}
    \centering
    \includegraphics[width=1.0\textwidth]{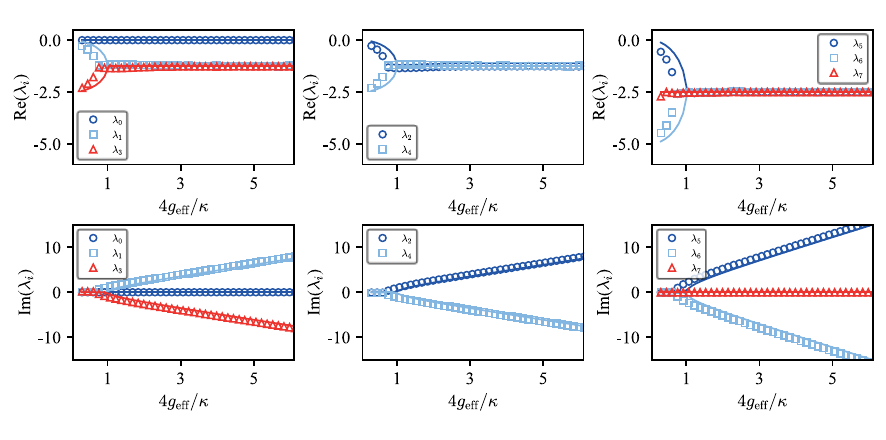}
    \caption{Numerically extracted Liouvillian spectrum with a fluctuation of the parametric modulation, $\delta\varepsilon=0.003\nu$. Here $g_{\rm eff}$ refers to the expected effective qubit-PM coupling strength without fluctuation}
    \label{fig3}
\end{figure}

Under the circumstances, the error function is defined as 
\begin{equation}
\text{Erf} = \vert A_j - C e^{Bt}\vert^2.
\end{equation}

The best-fit eigenvalues $\lambda_j$ are then the values minimizing the error function. The experimentally measured $A_j(t)$ (dots) and their fitting results (lines) are given in Fig.~\ref{fig1}.

\section{Effect of the fluctuation of the modulating amplitude}
The Hamiltonian describing the sideband interaction between the qubit (Q) and readout resonator (R) of our system is given by \cite{Phys.Rev.Lett.131.260201} 
\begin{equation}\label{H_sideband}
    H = g_r\left[\sum^{\infty}_{n=-\infty}J_n\left(\frac{\varepsilon}{\nu}\right)e^{-i(n\nu-\Delta_r)t}a^{\dagger}\vert g\rangle\langle e\vert + H.c. \right],
\end{equation}
where $g_r$ denotes the coupling strength between R and Q, $J_n(x)$ is the $n$-th Bessel function of the first kind, and $\Delta_r = \omega_r - \bar{\omega}$ is the detuning between the resonator frequency ($\omega_r/2\pi\simeq 6.60$ GHz) and the central frequency ($\bar{\omega}$) of the parametric modulation, whose amplitude and frequency are $\varepsilon$ and $\nu$, respectively. 

In our experiment, we tune the system parameters in such a way that ($\nu=\Delta_r$ and $g_r\ll\nu$) the first-order sideband modulation condition is satisfied. Under the circumstances, the off-resonant terms ($n\neq 1$) in the Eq.~(\ref{H_sideband}) can be neglected, reducing $H$ in Eq.~(\ref{H_sideband}) to 
\begin{equation}\label{geff}
    H = g_{\rm eff}a^{\dagger}\vert g\rangle\langle e\vert + H.c,
\end{equation}
where $g_{\rm eff} = g_rJ_1(\frac{\varepsilon}{\nu})$. The effective coupling strength $g_{\rm eff}$ can be controlled by tuning the modulation amplitude $\varepsilon$, with the first-order sideband condition satisfied by synchronously tuning the modulation frequency $\nu$.

To verify the validity of the effective Hamiltonian of Eq.~(\ref{geff}), we perform the simulation of qubit-PM evolution and extract the Liouvillian spectrum by fitting the output density matrix. Thus-obtained eigenenergies of the Liouvillian superoperator, versus the effective coupling strength, are displayed in Fig~\ref{fig2}, which agree well with the ideal results (lines), confirming that influences of the off-resonant terms in Eq.~(\ref{H_sideband}) are negligible. We now investigate the effect of the fluctuation of the modulation amplitude, denoted as $\delta\varepsilon$, which makes the effective qubit-PM coupling deviate from the expected value $g_{\rm eff}$. To quantify the sensitivity of the experimental results to such a fluctuation, we numerically simulate the system dynamics for $\delta\varepsilon=0.003\nu$ and extract the Liouvillian spectrum. The results, shown in Fig.~\ref{fig3}, show that the fluctuation only causes a slight deviation of the exceptional points.